\documentclass[conference]{IEEEtran}
\IEEEoverridecommandlockouts

\usepackage{cite}
\usepackage{amsmath,amssymb,amsfonts}
\usepackage{graphicx}
\usepackage{textcomp}
\usepackage{xcolor}
\usepackage{booktabs}
\usepackage{tikz}
\usepackage{url}
\usetikzlibrary{positioning,arrows.meta}

\begin{document}

\title{Revisiting Self-Attentive Sequential Recommendation Beyond the LLM Paradigm}

\author{
\IEEEauthorblockN{Zan Huang}
\IEEEauthorblockA{\textit{NVIDIA}\\
pehuang@nvidia.com}
}

\maketitle

\begin{abstract}
Sequential recommendation adopted the Transformer almost as soon as
it appeared: SASRec ported the decoder to next-item prediction in
2018, a year after \emph{Attention is All You Need}, and the paradigm
has borrowed from language modeling ever since.
The two tasks look nearly identical---both consume integer-ID
sequences with causal self-attention---yet they pursue opposite ends.
A recommender works to bring more users into contact with more items,
an entropy-\emph{increasing} goal; a language model works to converge
many phrasings of a question onto one answer, an
entropy-\emph{decreasing} one.
We argue this difference, not engineering effort, is why
recommendation has not reproduced the clean scaling that language
models enjoy: behavioral data is locally regular yet globally
heterogeneous---a \emph{casino}---whereas language is locally diverse
yet globally convergent---a \emph{library}.
Taking SASRec as an entry point, we revisit the self-attentive
paradigm as a \emph{comparative study} of the two domains and ask
which of its inherited assumptions---implicit-only personalization,
absolute positional semantics, leakage-prone single-step evaluation,
and atomic tokenization---are incidental rather than intrinsic to
recommendation.
Our BlueSky claim is that, beyond borrowing from language models, the
next findings will come from a careful comparison of the two domains
that starts from the entropy structure of behavioral data.
We propose no new model; we expose the gaps, outline the data- and
systems-level agenda they imply, and argue that the comparison can
ultimately help both domains.
\end{abstract}


\section{Introduction}

When Vaswani et al.~\cite{vaswani2017attention} published
\emph{Attention is All You Need} in 2017, the recommender-systems
community was quick to recognize its potential.
Within a year, SASRec~\cite{kang2018sasrec} adapted the Transformer
decoder to sequential recommendation, treating a user's interaction
history $(i_1, i_2, \ldots, i_T)$ as a token sequence and applying
causal self-attention to predict the next item.
The results were compelling: SASRec outperformed RNN-based
GRU4Rec~\cite{hidasi2016gru4rec} and CNN-based Caser~\cite{tang2018caser}
by clear margins on standard benchmarks, and its simplicity made it
an instant default baseline.

\begin{figure}[t]
\centering
\includegraphics[width=0.80\columnwidth]{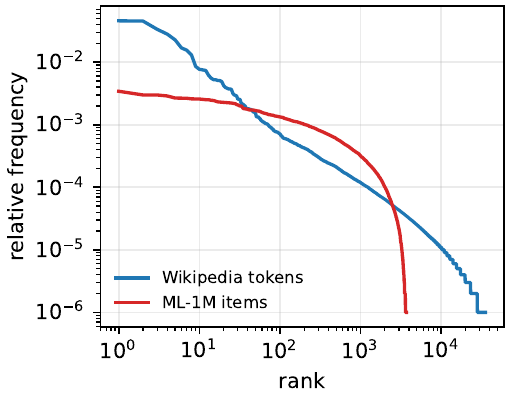}
\caption{Rank--frequency distributions (log--log) of symbol
occurrence: a deliberately coarse, order-invariant comparison of
marginal structure. \emph{ML-1M}: MovieLens-1M~\cite{harper2015movielens}
item IDs ($1{,}000{,}209$ interactions, $3{,}706$ items).
\emph{Wikipedia}: paragraphs from the \texttt{20220301.en} dump
tokenized with GPT-2 BPE~\cite{radford2019gpt2,sennrich2016bpe},
accumulated to a matched budget ($1{,}000{,}179$ tokens, $35{,}461$
distinct). Wikipedia closely follows a power law (top 1\% of the
vocabulary $\approx 52\%$ of the mass); ML-1M does not (top 1\%
$\approx 9\%$). Confounded by representation and vocabulary size, the
comparison is offered as motivation, not evidence.}
\label{fig:zipf}
\end{figure}

Beneath the syntactic resemblance, however, the two tasks pursue
opposite ends.
A recommender succeeds by bringing more users into contact with more
items, broadening and diversifying interaction; this is an
entropy-\emph{increasing} objective.
A language model succeeds by converging the many ways a question can be
asked onto one well-formed answer, an entropy-\emph{decreasing} one.
The difference shows up in the data.
A language corpus behaves like a \emph{library}: locally diverse in
surface form yet globally convergent in content, a regime of local
chaos with global order in which more capacity reliably extracts deeper
structure.
A stream of user behavior behaves like a \emph{casino}.
Any single history is sparse and repetitive, and each game looks
deceptively simple, yet the floor serves patrons whose tastes are
mutually incompatible: just as blackjack, poker, and slots share one
setting but reward entirely different strategies, users with orthogonal
preferences resist a single unified model, and the more data one
gathers the more this global heterogeneity compounds rather than
converges.
This is the opposite regime, local order with global chaos.
Unlike a language model, which learns from a fixed external corpus, a
recommender also sits inside a closed loop: its own past
recommendations shape the behavior that becomes its next training
signal, so each ``next item'' is less a free choice to be predicted
than a move in an ongoing game.
Figure~\ref{fig:zipf} makes the contrast concrete at a matched scale of
roughly one million symbols per corpus: Wikipedia tokens follow a clean
power law, the top 1\% of the vocabulary carrying roughly half the
probability mass, whereas ML-1M items have a far flatter, broken tail
whose top 1\% carries under one tenth.

Seen this way, the much-discussed observation that sequential
recommenders did not show language-model-style scaling is less a
modeling failure than a property of the data: recommendation \emph{has}
scaled, but largely \emph{out} (larger embedding tables, richer
features, longer histories~\cite{zhai2024hstu,chai2025longer}) rather
than \emph{up} along compute the way language models
do~\cite{kaplan2020scaling,hoffmann2022chinchilla}.

\emph{Why revisit the paradigm now?}
Because recommendation increasingly draws on language-model ideas, from
generative retrieval~\cite{rajput2023tiger} to LLM-based
recommenders~\cite{chen2024hllm}.
The relationship is, moreover, not one-directional.
Recommendation built much of the data and distribution infrastructure
on which large-scale language modeling now depends, and a growing
share of machine-generated content reaches an audience through
recommendation.
A comparison of the two is therefore in the interest of both: it is not
a question of which model wins a benchmark, but of which assumptions
transfer and which do not.

This is the BlueSky idea.
Taking SASRec, the canonical self-attentive recommender, as an entry
point, we treat it as the object of a \emph{comparative study} against
language modeling, and ask which of its inherited assumptions are
incidental rather than intrinsic to recommendation.
We do not propose a new model; we offer a systematic lens through which to examine issues that the field has been discussing in fragments.
Our claim is that the next durable progress will come from designing
around the intrinsic statistical structure of behavioral data, even as
the field continues to borrow productively from language models;
whether a good data power law is essential to a good model scaling law~\cite{brill2024neural}
is one of the questions we would pursue.
Section~\ref{sec:background} reviews the borrowed blueprint;
Section~\ref{sec:divergences} concentrates the inherited assumptions
worth questioning; Section~\ref{sec:realignments} sketches a
data-first agenda; and Section~\ref{sec:outlook} states what success
would look like and why both communities should care.

\section{Background}
\label{sec:background}

Before SASRec, sequential recommendation was dominated by two
paradigms: recurrent models such as GRU4Rec~\cite{hidasi2016gru4rec},
which processed histories token-by-token through gated recurrent
units, and convolutional models such as Caser~\cite{tang2018caser},
which applied learned filters over fixed windows of recent items.
Both had well-known limitations: RNNs from vanishing gradients over
long sequences, CNNs from rigid receptive fields that could not adapt
to variable-length dependencies.

SASRec addressed both at once.
By applying causal self-attention to the full sequence, it let each
position attend directly to any earlier one, capturing short- and
long-range dependencies in a single forward pass.
The architecture is clean: an item embedding table, a positional
embedding table, a stack of Transformer decoder blocks, and a
dot-product scoring head.
This simplicity made SASRec easy to implement, reproduce, and extend,
establishing it as the field's default starting point.
BERT4Rec~\cite{sun2019bert4rec} followed within a year with a
bidirectional, masked-item variant; later replication work attributes
much of its reported edge to the training objective rather than
bidirectionality
alone~\cite{petrov2023bert4rec,klenitskiy2023dross}.

Most subsequent advances continued in the same vein.
TiSASRec~\cite{li2020tisasrec} augments attention with explicit
positional \emph{and} inter-event time-interval biases, not time
alone; BSARec~\cite{shin2024bsarec} combines a Fourier-domain
inductive bias \emph{with} self-attention rather than replacing it;
SSE-PT~\cite{wu2020ssept} injects an explicit user embedding; a
separate thread redesigns the training loss to scale to large
catalogs~\cite{klenitskiy2023dross,mezentsev2024sce};
TIGER~\cite{rajput2023tiger} and GPTRec~\cite{petrov2023gptrec}
import tokenization ideas through semantic and sub-item IDs;
HSTU~\cite{zhai2024hstu} pushes the recipe to industrial scale; and
HLLM~\cite{chen2024hllm} folds pretrained language models in directly.

On the surface this reads as recommendation converging toward language
modeling, a trajectory often summarized as the rise of \emph{generative
recommendation}.
On closer reading, though, many of these advances diverge from the
language-model path, and several leave the Transformer's core largely
untouched, adding structure around it rather than scaling it up
aggressively.
The borrowing is thus better understood not as convergence but as
independent exploration under a shared name.
The next section takes up these divergences in turn.

\section{Divergences}
\label{sec:divergences}

\emph{Which inherited assumptions break, and why should the community
question them?}
We concentrate four, each a place where a language-shaped choice meets
casino-shaped data.

\subsection{Personalization That Never Enters Attention}

Academic benchmarks conventionally retain only user--item ID pairs,
discarding richer signals.
Under this setting, SASRec and its successors perform \emph{implicit}
personalization only: the input is filtered per user, but no explicit
user representation enters the computation.
Given an item embedding matrix
$\mathbf{E} \in \mathbb{R}^{|\mathcal{I}| \times d}$, a positional
matrix $\mathbf{P} \in \mathbb{R}^{T \times d}$, and a decoder
$f_\theta$, SASRec forms
$\mathbf{h} = f_\theta(\mathbf{E}_{i_1}{+}\mathbf{P}_1, \ldots,
\mathbf{E}_{i_T}{+}\mathbf{P}_T)$ and scores candidate $j$ as
$s(j)=\mathbf{h}^\top \mathbf{e}_j$.
User identity never enters.\footnote{See the reference implementations
\url{https://github.com/kang205/SASRec} (TensorFlow) and
\url{https://github.com/pmixer/SASRec.pytorch} (PyTorch): in both, the
user ID is read by the data sampler but never enters the model.}
Without negative sampling, two users with the same item sequence
receive identical scores; with it, the distinction is delegated to a
stochastic process.
The effect is sharpest for cold-start users, whose short,
near-identical histories give the model little else to tell them apart.

A curious duality exposes the deeper issue.
If a user can be represented by the items they touched, then by
symmetry an item can be represented by the users who touched it;
substituting one into the other yields an infinite regress, users
defined by items defined by users, with no ground-level anchor.
The way out is to stop treating the two symmetrically: items can be
anchored as fixed points, while a user is better described as a
trajectory over them, a different kind of object that the model never
explicitly forms.
Tellingly, many components added by follow-ups address the symptom
from outside the attention map rather than inside it: TiSASRec's joint
position-and-interval biases resemble the low-rank co-occurrence
structure of classical matrix factorization~\cite{koren2009matrix},
and even an explicit user model such as SSE-PT~\cite{wu2020ssept}
enters as an additive embedding rather than reshaping attention
itself.
The right question is therefore less ``is there a user vector'' than
``how much signal actually enters the attention computation versus
accumulating beside it'', and how much can be fused in.
At scale the same pattern recurs: recommenders deliver much of their
effectiveness through embedding tables and retrieval pipelines that sit
outside the attention and grow by scaling \emph{out}; even HSTU's
trillion-parameter scale~\cite{zhai2024hstu} resides overwhelmingly in
embedding tables and infrastructure rather than in attention depth.
The supervision compounds the problem: in deployed systems a user's
``next item'' was usually surfaced by a previous recommender, so the
label is conditioned on a prior policy---noisy, local, and time-bound.
The user's underlying interest is the ground truth, but it is
observable only through such bandit-style feedback, never directly.

\subsection{Position by Slot, Not by Target}

The original Transformer found learned and sinusoidal positional
encodings nearly equivalent~\cite{vaswani2017attention}; SASRec adopts
the learned variant.
The causal mask already distinguishes ``inside the prefix'' from
``outside''; it does not tell the model \emph{how far} each item is
from the prediction target; that is the positional embedding's job.
But SASRec assigns embeddings by absolute slot index
(Fig.~\ref{fig:posemb}a), so a given item carries the same positional
vector regardless of which next-item sub-task is active, and the
sub-tasks packed into one causal forward pass overlay incompatibly.
An aligned scheme (Fig.~\ref{fig:posemb}b) counts backward from the
target, so the immediate predecessor always carries the same code, but
realizing it requires restructuring training so each sub-sequence is
an independent example.

What the positional channel should carry is also worth reconsidering.
Picture a supermarket: items are goods at fixed coordinates in a
product space, and a user is a shopper tracing a \emph{path} through
it. The positional channel is the natural home for that path, encoding
the user's trajectory rather than a global slot ruler. Recommendation
nonetheless inherits from language the convention that positional and
item embeddings share a dimensionality, even though a trajectory may
have far lower intrinsic complexity than the item geometry, so a
lower-dimensional, decoupled positional channel is a degree of freedom
that language-inspired designs leave unused.
Relative schemes from language modeling, from the relative position
representations of Shaw et al.~\cite{shaw2018relative} to
RoPE~\cite{su2024rope} and ALiBi~\cite{press2022alibi}, along with
concurrent work~\cite{nabiev2026posaware,gusak2025split}, are starting
points, but the deeper mismatch between the positions the model encodes
and the ones the task needs sits upstream of any attention-level fix.

\begin{figure}[t]
\centering
\resizebox{0.58\columnwidth}{!}{%
\begin{tikzpicture}[
  ibox/.style={draw=gray!55, minimum width=1.55cm,
    minimum height=0.50cm, align=center, inner sep=1.5pt,
    rounded corners=2pt, font=\scriptsize\ttfamily},
  arr/.style={->, >=stealth, semithick},
]
\node[font=\footnotesize\bfseries, anchor=west]
  at (0.0, 1.05) {(a) Current: absolute slot indices};
\node[ibox] (a0) at (0.85, 0.45) {i[0]+p[-3]};
\node[ibox] (a1) at (2.50, 0.45) {i[1]+p[-2]};
\node[ibox] (a2) at (4.15, 0.45) {i[2]+p[-1]};
\draw[gray!25] (0.08, 0.18) -- (0.08,-0.58);
\draw[gray!25] (1.60, 0.18) -- (1.60,-0.24);
\draw[gray!25] (3.25, 0.18) -- (3.25,-0.41);
\draw[gray!25] (4.90, 0.18) -- (4.90,-0.58);
\draw[arr, blue!65]
  (0.08,-0.24) -- (1.60,-0.24) -- ++(0.30,0)
  node[right, font=\scriptsize, text=blue!65] {$\to i_1$};
\draw[arr, teal!65]
  (0.08,-0.41) -- (3.25,-0.41) -- ++(0.30,0)
  node[right, font=\scriptsize, text=teal!65] {$\to i_2$};
\draw[arr, purple!60]
  (0.08,-0.58) -- (4.90,-0.58) -- ++(0.30,0)
  node[right, font=\scriptsize, text=purple!60] {$\to i_3$};
\node[font=\scriptsize, text=red!55]
  at (2.9,-0.90)
  {all sub-tasks share p[-3], p[-2], p[-1]};
\begin{scope}[yshift=-2.2cm]
\node[font=\footnotesize\bfseries, anchor=west]
  at (0.0, 1.05) {(b) Aligned: distance-to-target};
\node[ibox] (r1a) at (0.85, 0.55) {i[0]+p[-1]};
\draw[arr, blue!65] (r1a.east) -- ++(0.30,0)
  node[right, font=\scriptsize, text=blue!65] {$\to i_1$};
\node[ibox] (r2a) at (0.85, 0.00) {i[0]+p[-2]};
\node[ibox] (r2b) at (2.50, 0.00) {i[1]+p[-1]};
\draw[arr, teal!65] (r2b.east) -- ++(0.30,0)
  node[right, font=\scriptsize, text=teal!65] {$\to i_2$};
\node[ibox] (r3a) at (0.85,-0.55) {i[0]+p[-3]};
\node[ibox] (r3b) at (2.50,-0.55) {i[1]+p[-2]};
\node[ibox] (r3c) at (4.15,-0.55) {i[2]+p[-1]};
\draw[arr, purple!60] (r3c.east) -- ++(0.30,0)
  node[right, font=\scriptsize, text=purple!60] {$\to i_3$};
\node[font=\scriptsize, text=teal!60]
  at (2.9,-0.95)
  {p[-1] always = immediate predecessor};
\end{scope}
\end{tikzpicture}%
}
\caption{Positional indexing in SASRec versus a target-aligned scheme.
A length-four history \texttt{(i[0],i[1],i[2],i[3])} is trained in one
causal pass that packs three next-item sub-tasks (predict
\texttt{i[1]}, then \texttt{i[2]}, then \texttt{i[3]}); \texttt{p[k]}
denotes the learned positional code at offset \texttt{k}, within a
window of \texttt{max\_len=3} positions.
(a)~SASRec indexes by \emph{absolute slot}, so all three sub-tasks
reuse \texttt{p[-3],p[-2],p[-1]} and overlay in one row, and an item's
code does not reflect its distance to the item being predicted.
(b)~A \emph{target-aligned} scheme indexes by distance to the target,
so the immediate predecessor always carries \texttt{p[-1]}; the cost is
that each sub-task must become a separate training example.}
\label{fig:posemb}
\end{figure}

\subsection{Evaluation That Hides the Gap}

The standard protocol (leave-last-out with sampled negatives) has
been questioned on several fronts: sampled metrics can be inconsistent
with exact ranking~\cite{krichene2020sampled}, well-tuned neural
models may not clearly beat simple baselines~\cite{dacrema2019progress},
the per-user leave-one-out split ignores the global timeline, letting
events that occur after a test point leak into
training~\cite{gusak2025split}, and several popular
benchmarks have weak sequential structure to begin
with~\cite{klenitskiy2026sequential}.
A single next-item hit can hardly separate a model that learned
preferences from one that approximates the marginal item distribution
rather than the user-conditioned one.
Crucially, a per-user split lets the future leak into the context: the
very construction that makes the task convenient also reproduces the
casino regime in miniature, rewarding the locally easy part while
hiding the globally hard part that blocks scaling.
A simple safeguard is to order interactions by global time before
splitting, so that no context contains an event later than the target
it predicts.

\subsection{Tokens Without Composition}

Language pipelines preprocess raw text with algorithms such as
BPE~\cite{sennrich2016bpe} that merge frequent character sequences
into composite tokens, shortening sequences and raising information
density.
Sequential recommendation long lacked an analogue: item IDs are atomic,
the vocabulary is the entire catalog, and the model must learn every
relationship from co-occurrence alone.
Semantic-ID and sub-item approaches such as TIGER~\cite{rajput2023tiger}
and GPTRec~\cite{petrov2023gptrec} take first steps toward such
tokenization; ActionPiece~\cite{hou2025actionpiece} goes further,
merging co-occurring feature patterns within and across adjacent
actions into context-dependent tokens.
We conjecture, rather than claim, that atomic item-ID sequences are
statistically thinner than BPE token sequences, with any given short
context recurring far less often,
as recommenders are designed to avoid repeating items;
testing this is itself an open problem.
Recommendation may ultimately need both directions: merging common
interaction patterns into composite units and decomposing rare items
into shared components.

These four problems are not independent; they are surface symptoms of
one underlying cause: behavioral data is structured very differently
from the text that language models consume. That is where the agenda
begins.

\section{Realignments}
\label{sec:realignments}

\emph{What are the unprecedented challenges, and where would we
start?} If the problems trace to how the data itself is structured, the
agenda is data-first.

\paragraph{Measure the data before scaling the model}
Treat the entropy structure of a sequence corpus as a measurable
workload: permutation entropy~\cite{bandt2002permutation},
compressibility, how fast the mutual information between two
interactions decays as the gap between them grows, and the slope of the
catalog's rank--frequency law (its Zipf exponent).
A clean way to isolate it is a controlled comparison: fix the
architecture and hyperparameters and vary only the data, contrasting
recommendation logs with language corpora cast into the same item-ID
form, so that any gap is attributable to the data rather than the
model.
Recommendation also affords data views that language lacks: the same
interaction log can be read as one sequence per user or, re-sorted by
global time, as one sequence per item. Training the \emph{same} SASRec
on each view yields markedly different accuracy
(Table~\ref{tab:multiview}), so what counts as ``the sequence'' is
 itself a modeling decision rather than a property of the data.


\begin{table}[t]
\centering
\caption{One fixed SASRec, two views of the same ML-1M user-item interactions: per user vs. per item, sorted by global time.}
\label{tab:multiview}
\small
\begin{tabular}{lcc}
\toprule
View & NDCG@10 & HR@10 \\
\midrule
ML-1M per-user & 0.592 & 0.821 \\
ML-1M per-item & 0.669 & 0.952 \\
\bottomrule
\end{tabular}
\end{table}

Whether a good data power law is essential to a good model scaling law
then becomes testable, with theoretical precedent: a line of work
derives a model's scaling exponent from the spectral or power-law
structure of its
data~\cite{sharma2022scaling,maloney2022solvable,bahri2024explaining,brill2024neural}.
Concurrent work fits recommender accuracy to data-quality terms such as
approximate entropy~\cite{shen2024optimizing}; our question is upstream
of such curve-fitting:
does a scaling power law for recommendation exist in the first place, and if so, what data property sets its exponent?

If the exponent is set by the data's spectrum, then the flat,
non-power-law tail of public recommendation catalogs
(Fig.~\ref{fig:zipf}) may be the mechanism itself rather than a
cosmetic detail, and diagnosing it must precede any scaling claim.
Entropy-based accuracy limits~\cite{ding2026entropy} are a
complementary tool, most informative once the distribution they
operate on has itself been diagnosed.
Defining such standard, information-theoretic workloads would be a
contribution in itself.

\paragraph{Separate scaling out from scaling up}
Name the two axes explicitly so that memory-scaling (embeddings,
features, retrieval) is not mistaken for the absence of
compute-scaling, and so that industrial claims of power-law
scaling~\cite{zhai2024hstu} are tested on public, reproducible
benchmarks rather than asserted from proprietary regimes.
That some large ranking systems are moving from pure self-attention
toward token-mixing and mixture-of-experts
designs~\cite{zhu2025rankmixer} is itself evidence that behavioral
heterogeneity resists a single attention-shaped model.

\paragraph{Fuse personalization into attention}
Make user identity and side information first-class inputs to the
attention computation rather than low-rank biases attached beside it,
and ask, for any mechanism, how much signal it actually fuses in.
A diagnostic is to track whether attention layers collapse toward the
effective rank of a co-occurrence matrix as depth or data grow.

\paragraph{Realign position and remove padding}
Encode distance-to-target rather than absolute slot, allow
domain-specific variants for temporal gaps and session boundaries, and
use a possibly lower-dimensional positional channel that carries the
user's trajectory through item space rather than a global slot ruler.
Realizing these changes also rewards leaner engineering. Processing
sequences at their natural, variable lengths rather than padding them
to a fixed size removes the artifacts that padding introduces, makes
target-aligned training practical, and uses the available compute more
fully, so that the algorithmic and engineering sides advance together.

\paragraph{Open a tokenization design space}
Explore merging frequent interaction patterns into composite tokens
and decomposing rare items into shared semantic components, extending
semantic-ID work~\cite{rajput2023tiger}.
Whether this changes the scaling characteristics of recommenders is an
open question.

\paragraph{Evaluate generatively and in order}
Autoregressive models can generate long continuations, yet we test a
single step.
Evaluating multi-step coherence in terms of diversity, entropy, and
alignment with held-out futures, while respecting global temporal
order, would surface failure modes that single-step protocols cannot,
such as mode collapse onto popular items or oscillation between
unrelated categories.
The generated sequences are themselves a lens on the data, a second
route to its entropy structure, and the most direct way to ask how a
recommender and a language model differ when each is left to continue a
sequence on its own.

\section{Outlook}
\label{sec:outlook}

SASRec showed that self-attention could be productively applied to
sequential recommendation and opened a research line that continues
eight years later.
Read with hindsight, its implicit-only personalization, absolute
positional semantics, leakage-prone evaluation, and atomic
tokenization are not independent quirks but interconnected signs that
a library-shaped architecture was carried onto casino-shaped data.

\emph{What would success look like, and why should both communities
care?}
Not another point on a leaderboard, but: a shared,
information-theoretic characterization of recommendation workloads;
architectures whose personalization and position genuinely live
inside attention; evaluation that respects temporal order and rewards
multi-step coherence; and a principled account of when and why
recommendation can scale.
Our contribution is diagnostic and agenda-setting rather than
prescriptive; we claim no benchmark gains.
Recommendation is among the largest deployed sequence-modeling
workloads on earth and supplies much of the data and distribution that
language models depend on; understanding where the two paradigms
diverge, and where they do not, is a question about data rather than
about models, and answering it stands to help both: recommendation
to grow more genuinely intelligent, and language models more genuinely
personalized.

\section*{Acknowledgment}
GenAI tools were used to improve readability.
The author is grateful to Srijan Kumar and Sejoon Oh for inspiring the 2020 PyTorch re-implementation of SASRec at Georgia Tech.

\bibliographystyle{IEEEtran}
\bibliography{revisiting_bluesky}

\end{document}